\documentclass[aps,prb,twocolumn,groupedaddress,showpacs,showkeys]{revtex4-1}

\usepackage{amsmath}
\usepackage{amssymb}
\usepackage[load-configurations=abbreviations,separate-uncertainty=true]{siunitx}
\usepackage{graphicx}

\usepackage{xcolor}
\newcommand{\sample}[1]{\emph{#1}}
\graphicspath{{figures/}}

\begin{document}

\title{Unusual anisotropic response of the charge carrier mobility to uniaxial mechanical strain in Rubrene crystals}

\author{Tobias~Morf}
\email[]{tmorf@phys.ethz.ch}

\author{Thomas~Mathis}
\email[]{mathis@phys.ethz.ch}


\author{Bertram~Batlogg}
\email[]{batlogg@phys.ethz.ch}
\affiliation{Laboratory for Solid State Physics, ETH Zurich, 8093 Zurich, Switzerland}

\date{\today}

\begin{abstract}
	Charge transport in Rubrene single crystals under uniaxial mechanical 
	strain is systematically investigated in the crystal's two in-plane 
	transport directions both under tensile and compressive strain applied 
	parallel or perpendicular to the current direction. 
	The density of trap states remains unchanged. 
	The field-effect mobility as a 
	benchmark figures for intermolecular transport is found to increase 
	with compressive strain and vice versa with a magnitude of 
	\SI{-1.5}{\cm\squared\per\V\per\s} per \si{\percent} of strain 
	independently of tranport direction. 
	A very remarkable result is the mobility change when the crystal is 
	strained perpendicular to the transport direction.
	While this enhancement could be quantitatively explained 
	from an improved wave-function overlap, mobility in the perpendicular 
	direction improves even more, contrary to simple geometric considerations 
	based od later expansion and usual Poisson ratios.
	This result emphasises the central role of the stress induced variations 
	of the dynamics wave function overlap in organic molecular crystals.
\end{abstract}


\maketitle

\section{Introduction}
	One of the main applications for organic semiconductors would be 
	flexible electronics. In such devices, the semiconductor will be 
	subject to mechanical strain and it is therefore paramount to 
	understand the effects of strain on the device properties. 
	Furthermore, this question is of fundamental interest for charge 
	transport in organic semiconductors. 
	The building blocks of an organic crystal are extended molecules 
	with richly structured electronic wave functions. These orbitals 
	are spatially highly structured including multiple nodes and lobes \cite{Bredas2002,daSilva2005,Coropceanu2007}. 
	Relatively small displacements are thus expected to cause drastic 
	modifications of the transfer integral and thus charge transport.
	
	Given such an intricate dependence on geometric modifications, it 
	appears highly desirable to study charge transport in single 
	crystals. Furthermore of interest are variations of transport 
	parallel and perpendiculare to the applied tensile and compressive 
	strain and along the principal crystallographic directions. Such 
	detailed measurements are an essential extension of previously 
	reported results \cite{Sekitani2005,Kanari2010,Someya2004,Lee2011,Chen2011,Sakai2013,Briseno2006}, 
	mainly on thin film devices which might also reveal effects of 
	grain boundaries and non-uniform crystallite orientation.

\section{Experiment}
\begin{figure}
		\centering
		\includegraphics[width=\columnwidth]{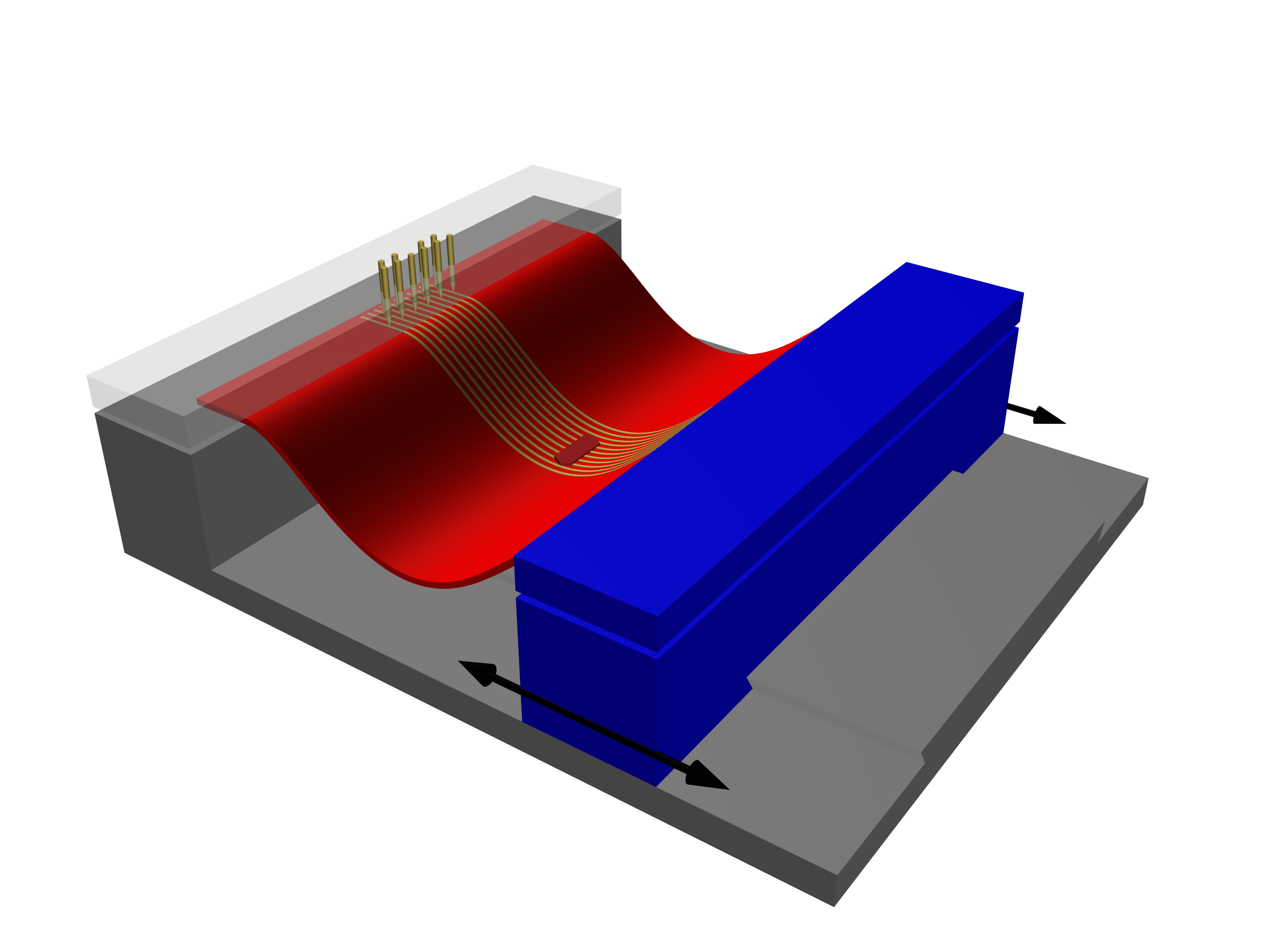}
		\caption{(Colour online) Illustration of the setup for transistor bending similar to a vice jaw. 
			The substrate (red) with prefabricated transistors is clamped to the jaws and the 
			bending radius set by adjusting the position of one jaw (blue). Electrical 
			contact is made through pins in the fixed jaw.\label{fig:setup}}
	\end{figure}

	Organic molecular crystals tend to be fragile and therefore precautions 
	are required when applying mechanical stress.
	Field-effect transistors were 
	therefore fabricated on flexible PEN foils \cite{Nakayama2013,Kato2004}, 
	\footnote{Teonex~Q65HA was kindly provided by Prof.~Jun Takeya's group at Chiba University} 
	and compressive or tensile strain exerted by bending the foil as shown in fig.~\ref{fig:setup}. 
	Strain at the interface is only a function of bending radius $R$ and amounts to $\varepsilon = \frac{t}{2R}$ 
	where $t$ is the foil thickness. 
	The electrodes, Cytop insulator \cite{Kalb2007} and \SI{\sim 1}{\um} thick Rubrene crystal 
	contribute with the third power of thickness to the 
	total bending moment \cite{Case1999,Suo1999,Chiang2009} and are thus neglible compared 
	to the $t=\SI{125}{\um}$ thick substrate. 
	
	For a continuous and controlled setting of the bending radius a vice-like 
	setup was built (fig.~\ref{fig:setup}). Here the 
	substrate is clamped to the jaws, with spring-loaded pins 
	on one jaw making the electrical connections. The distance 
	between the jaws was adjusted by a threaded rod.
	At well-defined bending radii corresponding to strain values between 
	\SI{-0.8}{\percent} (compressive) and \SI{+0.5}{\percent} (tensile) 
	field-effect transistor (FET) transfer characteristics were measured. 
	
	Since the wave function overlap depends on the intermolecular 
	distance and the field effect mobility $\mu$ is a measure of the overlap 
	it is thus expected to sensitively depend on the strain state 
	of the transistor. Apart from purely geometric considerations 
	mechanical strain may also influence the trap density of states (DOS) 
	through mechanical damage. This information can be obtained from 
	the subtheshold region of the transistor \cite{Blulle2014}. In the 
	following, we will therefore discuss the strain-related variations 
	of the mobility  in the linear regime $\mu_{\text{lin}}$ as well as the turn-on $V_{\text{on}}$ 
	and threshold voltage $V_{\text{th}}$.
	Since the turn-on voltage is characterised by the emerging from 
	the noise level and thus depends on the experimental setup, 
	and to reduce experimental scatter, $V_{\text{on}}$ was defined 
	at four different current levels well above the noise in the off-state 
	(\SIlist{10;30;100;300}{\pA}).
	A shift $\Delta V_{\text{on}}>0$ signifies 
	a shift to more positive voltages, i.e.\ to the right side 
	in fig.~\ref{fig:rawdata} where the $p$-type Rubrene transistor 
	is in the off state.

\section{Results and Discussion}
	
\begin{figure}
		\centering
		\includegraphics[width=\columnwidth]{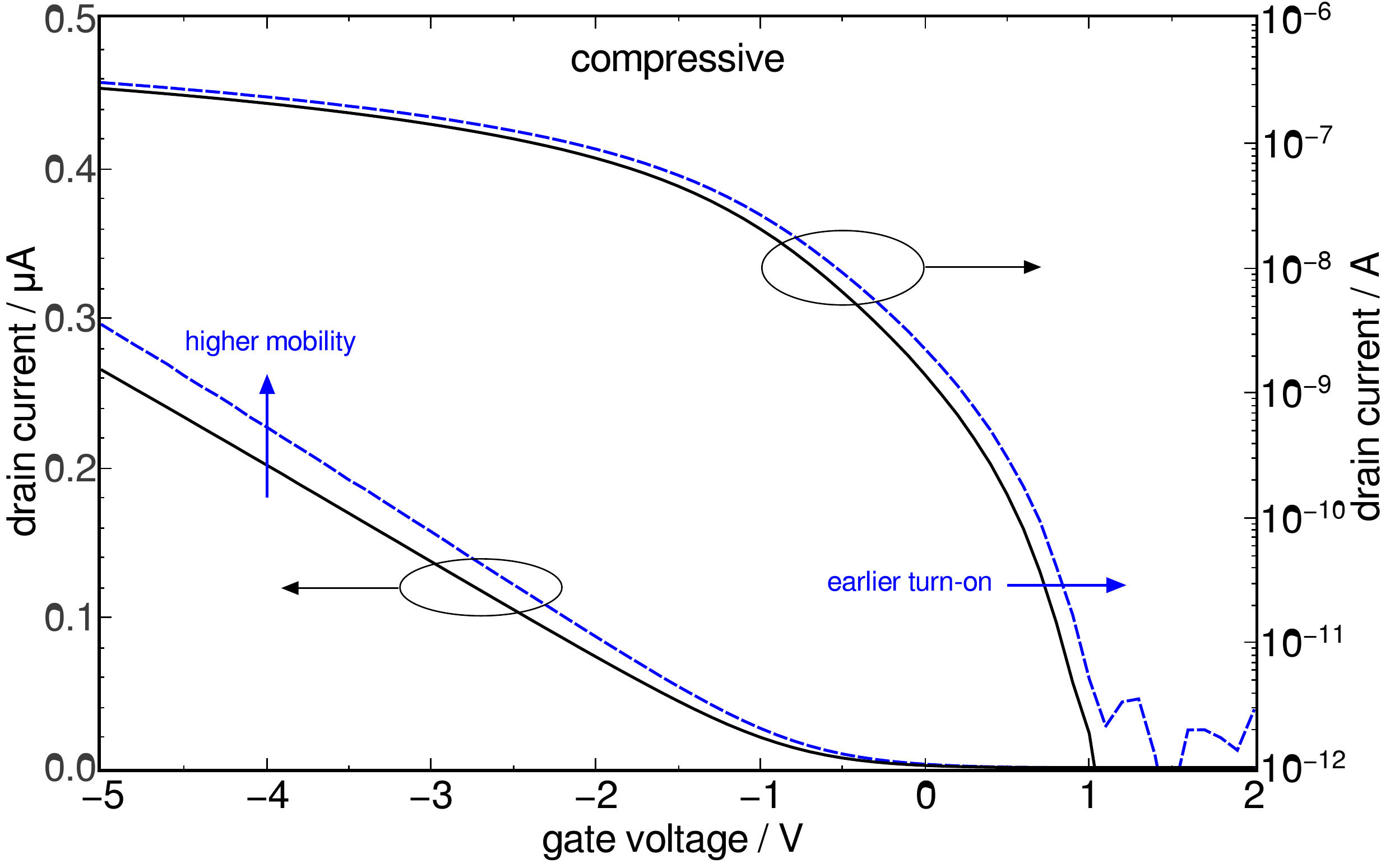}
		\includegraphics[width=\columnwidth]{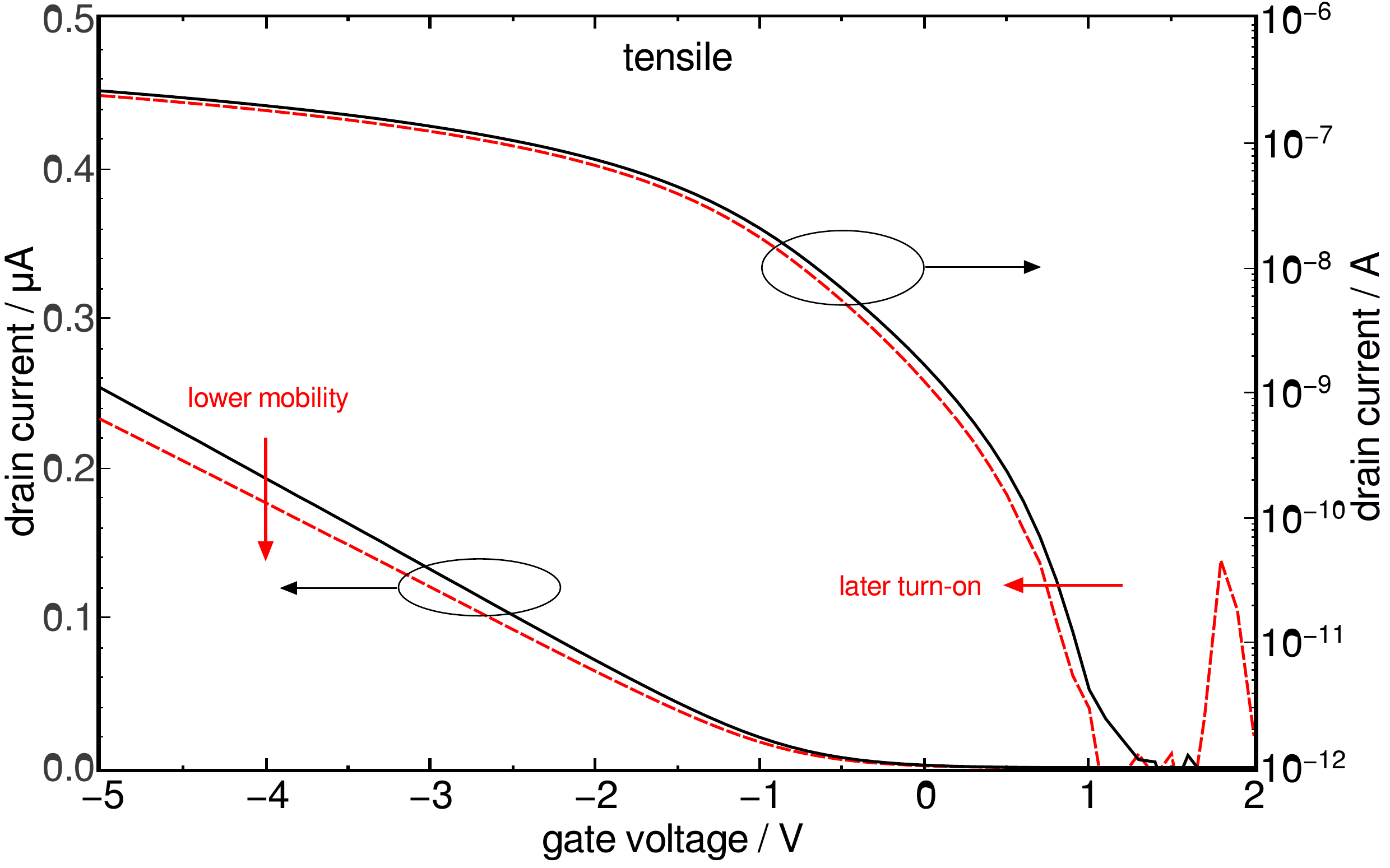}
		\caption{(Colour online) Transfer characteristics on a linear (left) 
			and logarithmic (right) scale. The black, solid curves represent 
			the relaxed state. Under compressive strain (top panel) the 
			mobility increases and the turn-on voltage shifts to the 
			right (blue, dashed lines). Vice versa for tensile strain 
			(bottom panel) where $\mu$ increases and $V_{\text{on}}$ 
			shifts to the left (red, dashed lines).\label{fig:rawdata}}
	\end{figure}

	For both the mobility and the turn-on voltage, the response to 
	externally applied strain usually consists of ai irreversible part 
	and a reversible one which is interpreted as damage and 
	elastic changes, respectively. While damage always results in 
	lower mobilities and a shift to the off-state the elastic 
	trend goes in both directions depending on the sign of the strain. 
	Fig.~\ref{fig:rawdata} shows the FET transfer characteristics: 
	mobility decreases with tensile strain and increases with compressive strain. 
	Similarly, the turn-on shifts to positive voltages with compressive
	strain and vice versa.
	
\begin{figure}
		\centering
		\includegraphics[width=\columnwidth]{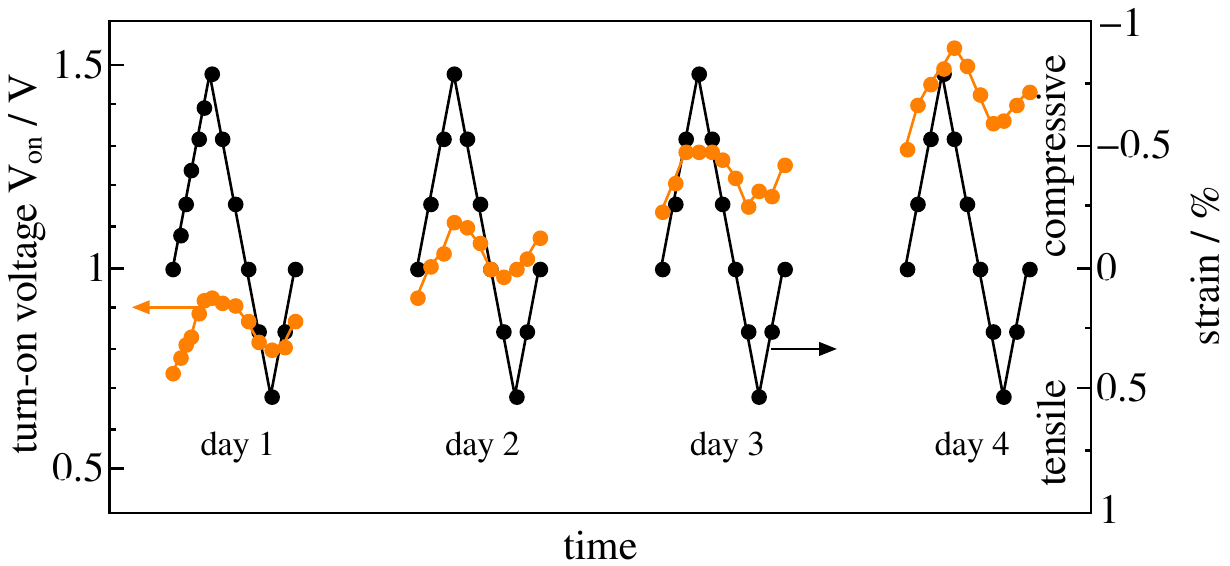}
		\caption{(Colour online) Typical measurement protocol (black) and corresponding turn-on 
			voltage (orange). One cycle consists of crystal compression, relaxation, 
			tension and relaxation again. At all times, there is a reversible shift 
			going parallel to this curve of applied strain. The overall trend 
			towards higher values of $V_{\text{on}}$ is interpreted as 
			damage.\label{fig:Vto_trend}}
	\end{figure}

\begin{figure}
		\centering
		\includegraphics[width=\columnwidth]{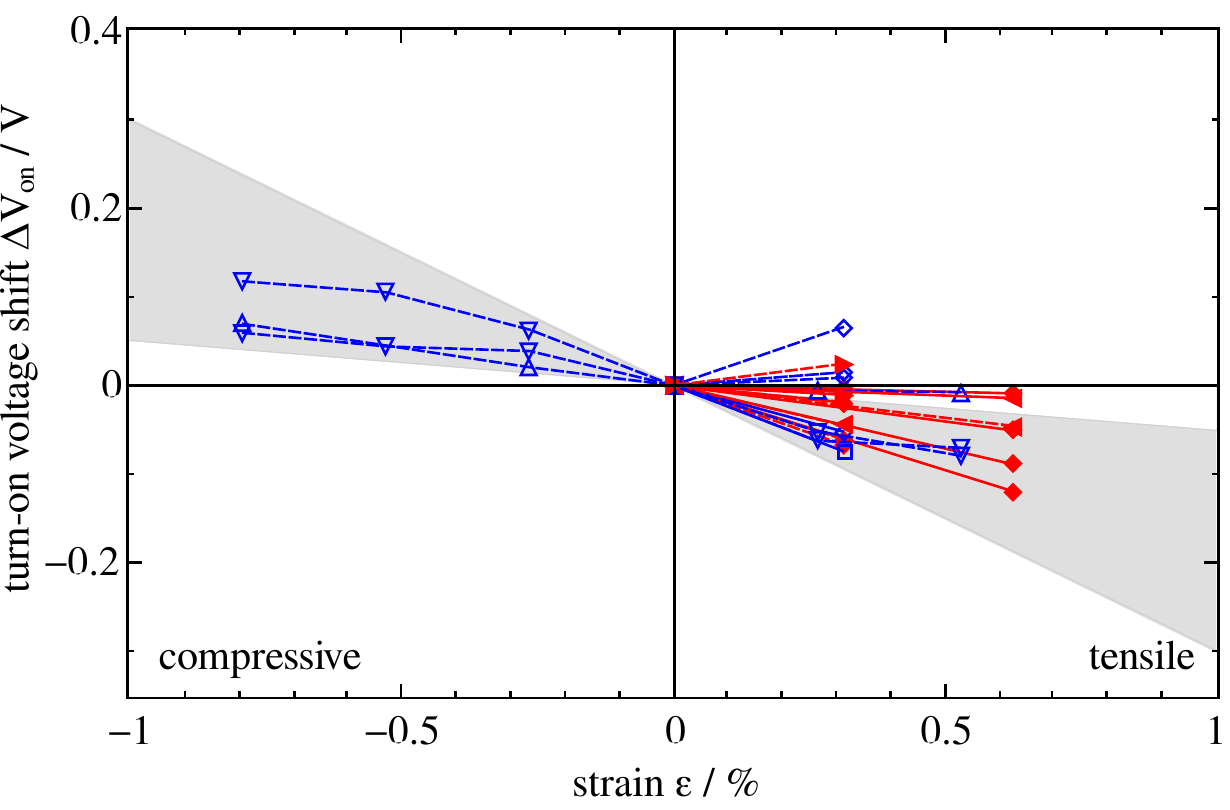}
		\caption{(Colour online) Summary of turn-on voltage vs.\ strain of all samples, coloured
			according to crystallographic transport direction. Blue: $a$-axis, red: $b$-axis transport. 
			Within the scatter of the data, there is not visible difference between the two 
			transport directions nor between parallel and perpendicular strain (solid and dashed 
			lines, respectively).\label{fig:Vto_master}}
	\end{figure}

	The measurement protocol involves in general several cycles of 
	compressive and tensile strain over the course several days.
	Turn-on voltages and corresponding strain values for a typical 
	sample are shown as a timeline in fig.~\ref{fig:Vto_trend}.
	The transistors' turn-on voltage reacts to the application and 
	relaxation of strain in a largely reversible way by shifting 
	\SI{\approx -0.2}{\V\per\percent} and it does so reproducibly and 
	repeatedly over several days in addition to a small, non-recovering 
	shift ascribed to damage. This trend is generally observed in 
	all the crystals shown in fig.~\ref{fig:Vto_master}.
	
	We may express the $V_{\text{on}}$ shift in terms of additional 
	charges at the semiconductor-dielectric interface because the 
	shift associated with compression does not 
	only move the turn-on in the direction of the off-state, but 
	actually into the positive gate voltage range. 
	A shift of \SI{0.2}{\V} at 
	\SI{-1}{\percent} (compressive) strain corresponds to an 
	interface charge density of $\SI{0.2}{\V}\cdot\SI{3.7}{\nano\farad\per\square\cm}/e = \SI{5e9}{\per\square\cm}$. 
	With the unit cell parameters $a=\SI{14.24}{\angstrom}$, $b=\SI{7.17}{\angstrom}$
	and two layers of two molecules per unit cell, this translates 
	to two new holes in \num{e5} interface molecules.
	
\begin{figure}
		\centering
		\includegraphics[width=\columnwidth]{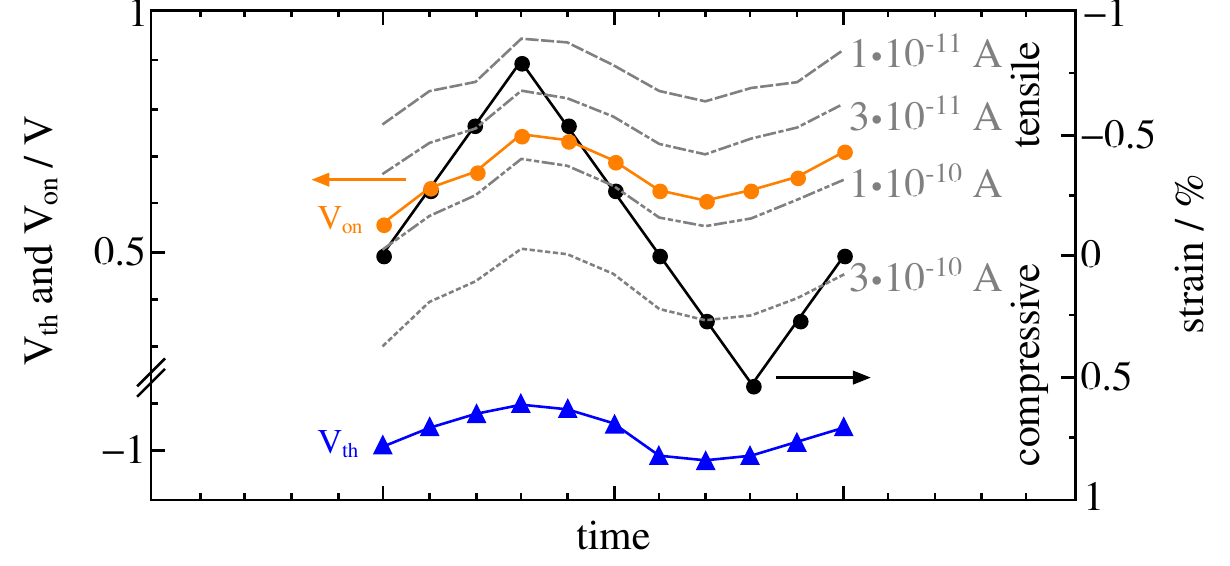}
		\caption{(Colour online) Threshold and turn-on voltage as a function 
			of applied strain. The four gray curves for $V_{\text{on}}$ are 
			averaged to give the orange one. All curves are parallel 
			meaning that the entire subthreshold region is 
			shifted uniformly without affecting the subthreshold slope. 
			This indicates that no new deep trap states are created 
			by mechanical strain.\label{fig:Vth_vs_Vto}}
	\end{figure}

	Remarkably, the shift of $V_{\text{on}}$ follows the sign change of 
	the strain, suggesting it to be not simply caused by a mechanical 
	modification of the crystal surface. Upon closer inspection, the 
	$V_{\text{on}}$ shift reflects a shift of the entire transfer curve. 
	In fig.~\ref{fig:Vth_vs_Vto} $\Delta V_{\text{on}}$ and $\Delta V_{\text{th}}$ 
	are shown to follow a strain cycle quantitatively in the same way.
	The threshold voltage $V_{\text{th}}$, the turn-on voltage $V_{\text{on}}$
	as well as the four defining voltages for $V_{\text{on}}$ are shifted 
	by the same amount. This means that 
	the entire subthreshold region is shifted and thus the deep trap 
	density \cite{Blulle2014} remains unchanged on the level of \SI{e15}{\per\cubic\cm\per\eV}.
	
\begin{figure}
		\centering
		\includegraphics[width=\columnwidth]{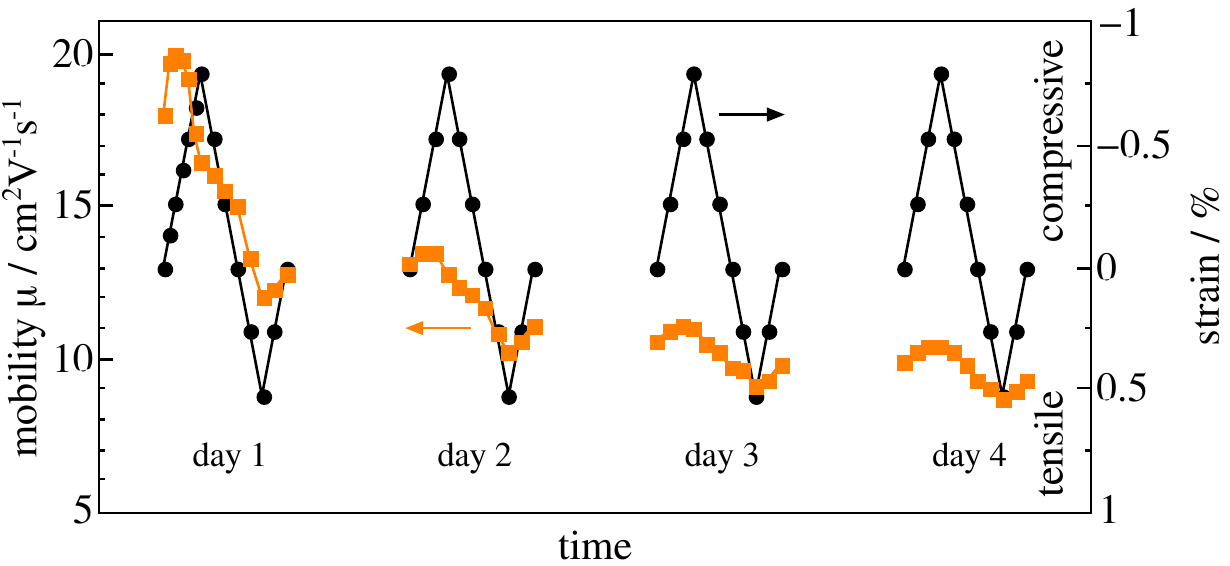}
		\caption{(Colour online) Typical measurement protocol (black) and corresponding mobility 
			values (orange) on four consecutive days. As for the turn-on 
			voltage, the mobility follows the strain curve in a revsible way 
			indicating an elastic response of the crystal to strain with 
			an associated variation of the molecular orbital overlap.
			The decreasing trend is attributed to non-elastic, irreversible damage, 
			i.e.\ cracks effectively reducing the FET channel geometry.\label{fig:mu_protocol}}
	\end{figure}

\begin{figure}
		\centering
		\includegraphics[width=\columnwidth]{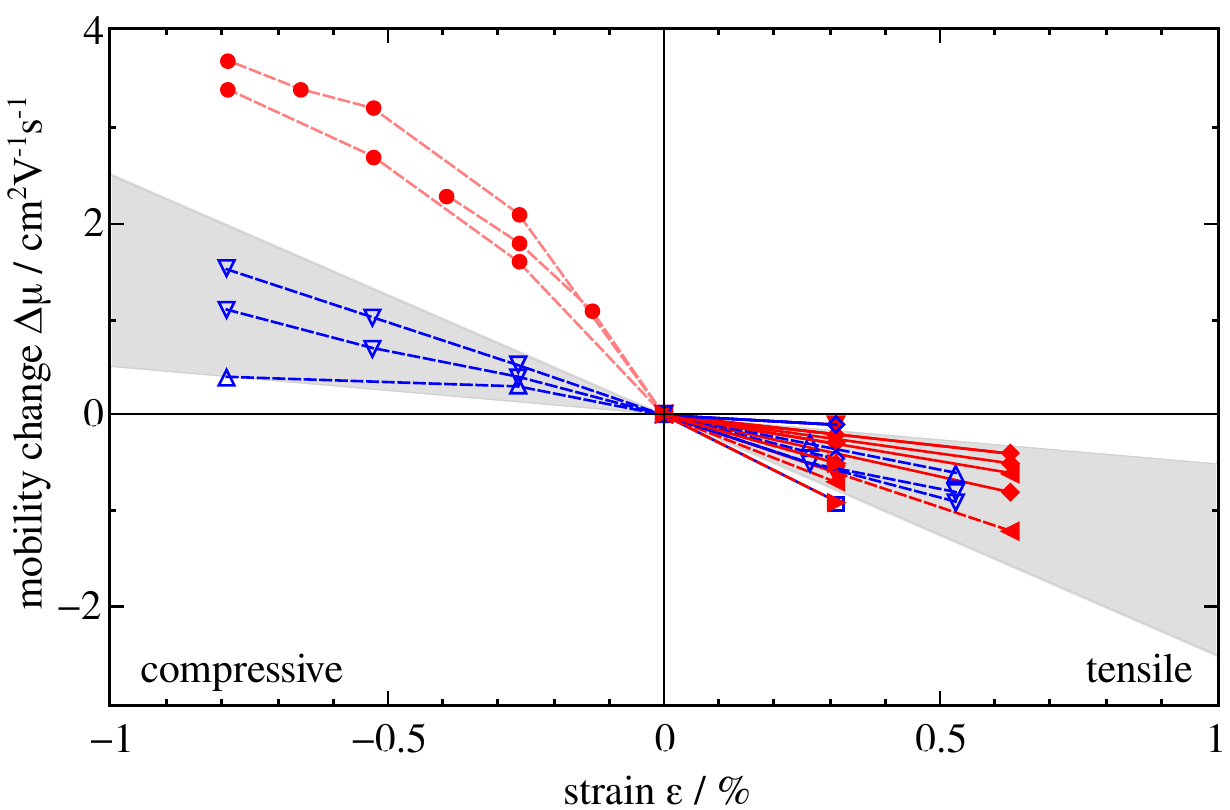}
		\caption{(Colour online) Summary of mobility vs.\ strain of all samples, coloured
			according to crystallographic transport direction as in fig.~\ref{fig:Vto_master}. 
			The general trend for the elastic response is \SI{\approx -1.5}{\cm\squared\per\V\per\s\per\percent}. 
			One crystal lies well above this range and is therefore considered 
			the upper limit for elastic behaviour due to the absence of visible cracks in the crystal.\label{fig:mu_master}}
	\end{figure}

	Similarly surprising are the strain effects on the charge dynamics, 
	measured in terms of the hole mobility $\mu_{\text{lin}}$.
	The timeline is again shown for four complete 
	strain cycles in fig.~\ref{fig:mu_protocol}. 
	The mobility always follows the curve of the applied strain and 
	we observe visible cracks accounting for a reduction of the effective 
	channel width and hence for the overall trend to lower values of 
	the apparent mobility.
	The changes $\Delta\mu/\varepsilon$ do not depend on the initial value 
	of $\mu_{\text{lin}}$ and amount to approximatly \SI{-1.5}{\cm\squared\per\V\per\s\per\percent} 
	as shown in fig.~\ref{fig:mu_master} on nine transistors.
	Within the scatter of the data, the crystal's $a$- and $b$-directions 
	respond similarly.
	
	A notable exception to the general trend is transistor \sample{1402s} in 
	fig.~\ref{fig:mu_master}. With $\Delta\mu/\epsilon \approx \SI{-5.6}{\cm\squared\per\V\per\s\per\percent}$ 
	it reacts much stronger to the applied strain. Despite this 
	being only one sample, the data are considered very reliable since the 
	transitor is of high quality and the characteristics are very reproducible 
	over multiple strain cycles. 
	The high mobility in the unstrained state of \SI{25}{\cm\squared\per\V\per\s} 
	remaines constant within less than \SI{1}{\cm\squared\per\V\per\s} and the crystal 
	does not show any visible cracks in the FET channel area.
	The very pronounced reaction to mechanical strain can thus be understood 
	as an upper limit to the purely elastic response.
	
	The over-all relatively large influence of strain on the mobility 
	reflects the modification of wave function overlap due to modified 
	relative geometric arrangement and of the dynamics of the molecule. 
	A first, over-simplified estimate would only consider compression 
	of the lattice, i.e. bringing the molecules closer together without 
	any additional slipping or rotating. Such an estimate would result 
	in an increase of the overlap by \SI{\approx 10}{\percent} per \SI{1}{\percent} 
	distance reduction \cite{Bredas2002} and is consistent with 
	the above measured value of $\Delta\mu/\varepsilon \approx \SI{-1.5}{\cm\squared\per\V\per\s\per\percent}$ 
	for both compressive and tensile strain.
	
\begin{figure}
		\centering
		\includegraphics[width=\columnwidth]{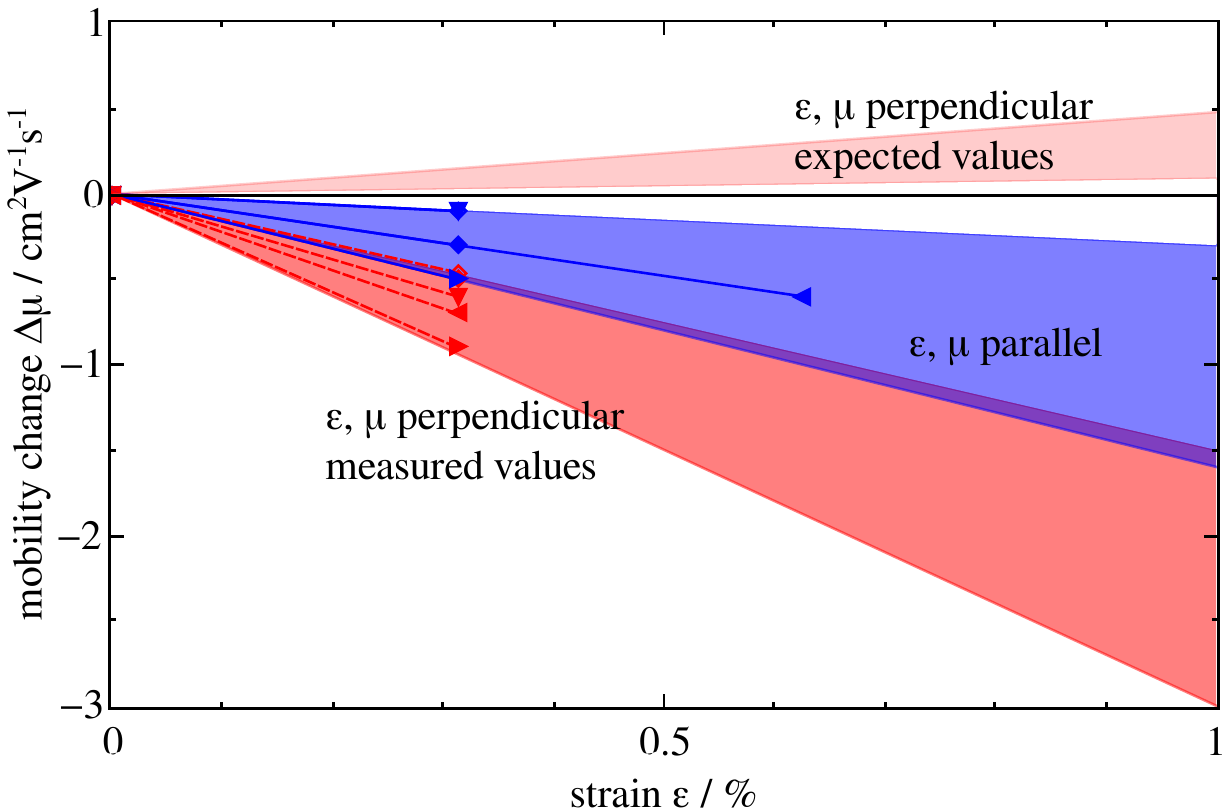}
		\caption{(Colour online) Mobility change $\Delta\mu/\epsilon$ for five 
			crystals measured with strain parallel to the transport 
			direction (blue) and perpendicular (red). 
			Contrary to the expectation from Poisson's law (light red wedge above $\Delta\mu=0$, 
			the absolute values are higher in the perpendicular configuration 
			and they have the same sign as the parallel ones.\label{fig:dmu_s-p}}
	\end{figure}

	Unexpected and unusual effects are observed then current flows 
	perpendicular to the strain direction. To eliminate extraneous 
	effects, the same transistor was measured in such a way that the 
	strain is either parallel or perpendicular to the current 
	direction. The mobility change of four crystals with current along 
	the $b$-axis and one along $a$ is shown in fig.~\ref{fig:dmu_s-p}. 
	In a simple situation, considering only macroscopic geometry, 
	due to lateral contraction and Poisson ratios $\nu\approx\num{0.3}$ 
	one would expect the mobility change with strain perpendicular 
	to transport (top, light red area) to be of opposite sign and lower magnitude 
	compared to the parallel case (blue area). Most remarkably 
	however, in Rubrene, in the perpendicular configuration the 
	mobility response is of the same sign and even larger (bottom, red area) 
	than in the parallel setting. Again, $a$- and $b$-axis transport 
	respond similarly.
	
	This unusual response is yet another example of 
	unexpected and nontrivial, anisotropic behaviour of organic 
	molecular crystals after e.g.\ the observation of thermal contraction 
	in Pentacene \cite{Haas2007}. While thermal expansion/contraction is 
	a strictly strucural effect, our data reveal that 
	the consequences extend to the electronic properties as well.
	
\section{Conclusion}
	In a systematic study of Rubrene crystal FETs the charge transport 
	and transfer characteristics have been investigated in all possible 
	parameter combinations: compressive and tensile strain along the 
	two principal crystal directions and current flowing parallel or 
	perpendicular to the strain.
	
	We measured a systematic reversible shift of the essentially unchanged transfer 
	curve by \SI{-0.2}{\V\per\percent} corresponding to two new holes 
	per \num{e5} interface molecules. This shift is symmetric for 
	tensile and compressive strain and no dependence on the transport 
	direction in the crystal was found. It suggests only minimal 
	deep trap creation but variations in the effective potential at 
	the semiconductor-dielectric interface.
	
	The charge carrier mobility responds in an unexpected and anisotropic 
	way to the applied strain. The general trend of \SI{-1.5}{\cm\squared\per\V\per\s\per\percent} 
	is also symmetric, independent of the transport direction and in line 
	with simple estimates based on transfer integral variations upon 
	intermolecular distance change. However, the difference between strain 
	applied parallel and perpendicular to current is most remarkable. 
	If tensile strain is applied perpendicular 
	to the transport direction, the crystal's mobility drops and it 
	does so even stronger than for parallel tensile strain.
	Both qualitatively and quantitatively, this is at odds with simple 
	geometrical considerations involving lateral contraction.
	
	This is an  example of unexpected, anisotropic behaviour of 
	organic single crystals reminiscent of negative thermal expansion 
	in Pentacene single crystals\cite{Haas2007}. Apart from macroscopic geometric 
	effects, the electronic structure is particularly sensitive to the relative 
	alignment of the molecules due to their highly structured wave 
	functions and the dynamic modulation of their overlap\cite{Bredas2002,Fratini2016,Troisi2006,Troisi2006b,Troisi2007,Sleigh2009}.
	In a unique way 
	the charge mobility is determined by the subtle interplay between 
	quantum localisation on a short timescale due to electronic 
	disorder (both of diagonal and mainly off-diagonal nature) and 
	the subsequent diffusive delocalisation driven by lattice dynamics 
	(for an updated summary, see e.g.\ ref.~\onlinecite{Fratini2016}).
	
	Thus our results stimulate a multitude of further detailed studies 
	of geometry and dynamics of the molecular arrangement under various 
	stress conditions. This will be a challenge for computational and 
	experimental approaches.
	
	\begin{acknowledgments}
	Kurt Mattenberger is greatfully acknowledged for technical support.
	\end{acknowledgments}
	
	\bibliography{../../../Papers-Bib/Library_Full,../../../Papers-Bib/Library_unread}

\end{document}